\documentstyle[12pt]{article}
\makeatletter
\def\bigans{y }
\read-1 to\answ
\ifx\answ\bigans
\ifcase\@ptsize
 \font\tenmsa=msam10
 \font\sevenmsa=msam7
 \font\fivemsa=msam5
 \font\tenmsb=msbm10
 \font\sevenmsb=msbm7
 \font\fivemsb=msbm5
\or
 \font\tenmsa=msam10 scaled \magstephalf
 \font\sevenmsa=msam8
 \font\fivemsa=msam6
 \font\tenmsb=msbm10 scaled \magstephalf
 \font\sevenmsb=msbm8
 \font\fivemsb=msbm6
\or
 \font\tenmsa=msam10 scaled \magstep1
 \font\sevenmsa=msam8
 \font\fivemsa=msam6
 \font\tenmsb=msbm10 scaled \magstep1
 \font\sevenmsb=msbm8
 \font\fivemsb=msbm6
\fi

\newfam\msafam
\newfam\msbfam
\textfont\msafam=\tenmsa  \scriptfont\msafam=\sevenmsa
  \scriptscriptfont\msafam=\fivemsa
\textfont\msbfam=\tenmsb  \scriptfont\msbfam=\sevenmsb
  \scriptscriptfont\msbfam=\fivemsb

\def\hexnumber@#1{\ifnum#1<10 \number#1\else
 \ifnum#1=10 A\else\ifnum#1=11 B\else\ifnum#1=12 C\else
 \ifnum#1=13 D\else\ifnum#1=14 E\else\ifnum#1=15 F\fi\fi\fi\fi\fi\fi\fi}

\def\msa@{\hexnumber@\msafam}
\def\msb@{\hexnumber@\msbfam}

\catcode`\@=12
\def\ZZ{\Bbb Z}
\def\CC{\Bbb C}

\else
\def\CC{I\!\!\!\!C}
\def\ZZ{Z\!\!\!Z}
		\fi

\def\tR{{\widetilde R}}
\def\ex{{\rm e}}
\def\sign{\eta}
\def\sig{\sigma}
\def\om{\omega_\sig^{\G,r}}
\def\gauge{ e}
\def\W{\Upsilon}
\def\row{\ell}
\def\col{k}
\def\G{G}
\def\Eu{{\cal E} }
\def\hchi{{\raise 0.8mm \hbox{$\chi$}}}
\def\cA{{\cal A}}
\def\cM{{\cal M}}
\def\cO{{\cal O}}

\def\cY{{\cal Y}}
\def\sun{{\rm SU}(N)}
\def\spn{{\rm Sp}(N)}
\def\son{{\rm SO}(N)}
\def\sinS{\sig \in S_r}
\def\RinY{R \in \cY_r}

\def\sst{\scriptscriptstyle}
\def\sR{{\sst R}}

\def\ds{\displaystyle}
\def\ts{\textstyle}

\catcode`@=11
\def\citen#1{\if@filesw \immediate\write \@auxout {\string\citation{#1}}\fi%
\@tempcntb\m@ne \let\@h@ld\relax \def\@citea{}%
\@for \@citeb:=#1\do {\@ifundefined {b@\@citeb}%
    {\@h@ld\@citea\@tempcntb\m@ne{\bf ?}%
    \@warning {Citation `\@citeb ' on page \thepage \space its cite}}%
    {\@tempcnta\@tempcntb \advance\@tempcnta\@ne
    \setbox\z@\hbox\bgroup\ifcat0\csname b@\@citeb \endcsname \relax
    \egroup \@tempcntb\number\csname b@\@citeb \endcsname \relax
    \else \egroup \@tempcntb\m@ne \fi \ifnum\@tempcnta=\@tempcntb
    \ifx\@h@ld\relax \edef \@h@ld{\@citea\csname b@\@citeb\endcsname}%
    \else \edef\@h@ld{\hbox{--}\penalty\@highpenalty
    \csname b@\@citeb\endcsname}\fi
    \else \@h@ld\@citea\csname b@\@citeb \endcsname \let\@h@ld\relax \fi}%
\def\@citea{,\penalty\@highpenalty\hskip.13em plus.13em minus.13em}}\@h@ld}
\def\@citex[#1]#2{\@cite{\citen{#2}}{#1}}%
\def\@cite#1#2{\leavevmode\unskip\ifnum\lastpenalty=\z@\penalty\@highpenalty\fi%
   $^{\scriptscriptstyle \multiply\@highpenalty 3 \mbox{\rm\scriptsize#1%
  \if@tempswa,\penalty\@highpenalty\ #2\fi}}$}   %

\def\dciten#1{\if@filesw \immediate\write \@auxout {\string\citation{#1}}\fi%
\@tempcntb\m@ne \let\@h@ld\relax \def\@dcitea{}%
\@for \@dciteb:=#1\do {\@ifundefined {b@\@dciteb}%
    {\@h@ld\@dcitea\@tempcntb\m@ne{\bf ?}%
    \@warning {line Citation `\@dciteb ' on page \thepage \space its dcite}}%
    {\@tempcnta\@tempcntb \advance\@tempcnta\@ne
    \setbox\z@\hbox\bgroup\ifcat0\csname b@\@dciteb \endcsname \relax
    \egroup \@tempcntb\number\csname b@\@dciteb \endcsname \relax
    \else \egroup \@tempcntb\m@ne \fi \ifnum\@tempcnta=\@tempcntb
    \ifx\@h@ld\relax \edef \@h@ld{\@dcitea\csname b@\@dciteb\endcsname}%
    \else \edef\@h@ld{\hbox{--}\penalty\@highpenalty
    \csname b@\@dciteb\endcsname}\fi
    \else \@h@ld\@dcitea\csname b@\@dciteb \endcsname \let\@h@ld\relax \fi}%
\def\@dcitea{,\penalty\@highpenalty\hskip.13em plus.13em minus.13em}}\@h@ld}
\def\@dcitex[#1]#2{\@dcite{\dciten{#2}}{#1}}%
\def\@dcite#1#2{\leavevmode\ifnum\lastpenalty=\z@\penalty%
  \@highpenalty\fi%
 {\multiply\@highpenalty 3
   {\rm [#1]\if@tempswa,\penalty\@highpenalty  #2\fi}}}   %
\def\dcite{\@ifnextchar [{\@tempswatrue\@dcitex}{\@tempswafalse\@dcitex[]}}
\catcode`@=12

\newcommand{\eq}{\begin{equation}}
\newcommand{\en}{\end{equation}}
\newcommand{\ie}{{\it i.e.}}
\def\hep#1#2#3#4{hepth$\bullet$}
\def\smb{{\footnotesize $\bullet$}}

\begin{document}
\setlength{\unitlength}{0.25cm}
\thispagestyle{empty}

\hfill              \begin{tabular}{l} {\bf \hep-th/9310105} \\
                                        {\sf BRX-TH--352} \\
                                        {\sf JHU-TIPAC--930024} \\
                                        {\sf BOW-PH--101} \\
                                    \end{tabular}

\vspace{1.7cm}

\begin{center}
\begin{tabular}{c}
{\LARGE \kern-26pt {Twist Points as Branch Points for the QCD$_2$ String}}
\end{tabular}
\vspace{1.5cm}

\setcounter{footnote}{1}
{\large Stephen G. Naculich,\footnote{Supported in part by
the NSF under grant PHY-90-96198
and by the Texas National Research Laboratory Commission
\setcounter{footnote}{3}
under grant RGFY-93-292}\footnote{Present address}
\setcounter{footnote}{2}
 Harold A. Riggs,\footnote{Supported in part by  the DOE under grant
               DE-FG02-92ER40706}
 and Howard J. Schnitzer$^\ddagger$}
\vspace{0.5cm}
{ \normalsize \sl
\begin{tabular}{lcl}
\begin{tabular}{c}
$^{\dagger \mathchar"278}$Dept. of
           Physics and Astronomy \\
    Bowdoin College \\
 Brunswick, ME  04011
\end{tabular}    &  {\it and} &
\begin{tabular}{c}
$^{\dagger}$Dept. of Physics and Astronomy \\
    The Johns Hopkins University  \\
 Baltimore, MD  21218
\end{tabular}                             \\
\multicolumn{3}{c}{naculich@polar.bowdoin.edu}
\end{tabular}    }
\vspace{0.5cm}

{ \normalsize \sl
\begin{tabular}{c}
                          $^\ddagger$Department of Physics \\
                                 Brandeis University  \\
                               Waltham, MA 02254    \\[0.1cm]
    \begin{tabular}{r}
          hriggs \\
         schnitzer
   \end{tabular}
    \kern -0.9em { @binah.cc.brandeis.edu}
\end{tabular} }
\end{center}

\vfill
\begin{center}
{\sc Abstract}
\end{center}

\begin{quotation}

We show that the string representation of the QCD$_2$ partition function
satisfies, by virtue of a Young-tableau-transposition symmetry, the
topological constraint that any branched covering of
an orientable or nonorientable surface without boundary
must have an {\em even} branch point
multiplicity.  This statement holds for each chiral sector and requires
multiple branch point behavior for the twist points, since cross-terms
appear that couple twist points with odd powers of simple branch points.
We obtain the same result for the
complete partition function of $\son$ and $\spn$
Yang-Mills$_2$ theory.
\end{quotation}
\vfill
{\sf October 1993} \hfill

\setcounter{page}{0}
\newpage
\setcounter{page}{1}

The long-anticipated connection between large-$N$ QCD and string theory
has recently been made precise for two-dimensional Yang-Mills theory
with gauge groups $\sun$, $\son$,
and $\spn$.\cite{gross,minahan,grotay,twist,dk,nrs,ram,kos,recent}
In particular, it has been shown that the $1/N$ expansion
of the Yang-Mills partition function can be interpreted geometrically
as a sum over maps from the string worldsheet onto the
(Euclidean) spacetime $\cM$ on which the Yang-Mills theory is
formulated. The leading terms in the $1/N$ expansion
count topologically distinct (unbranched) coverings of $\cM$
by the string worldsheet.
The primary subleading terms correspond to coverings with
simple
branch points as well as to (branched and unbranched) coverings that
map infinitesimal handles (for $\sun$)
or infinitesimal cross-caps (for $\son$ and $\spn$) on the worldsheet
to points on $\cM$. When $\cM$ is not a torus or Klein bottle
(so that the Euler characteristic $\Eu$ of $\cM$ is nonzero),
additional subleading terms arise
that have properties reminiscent of coverings with multiple branch points.
While these so-called {\it twist-point} terms  do
produce permutations of the sheets of the covering,
as would be expected if the twist points were multiple branch points, they
have another property that raises a puzzle
for this interpretation. Namely, while each simple branch point term
appears with a single power of the area of $\cM$,
the twist point terms do not carry any
powers of the area
(despite the fact that multiple branch points
are homotopically equivalent to a confluence of simple branch points).
This has led to the speculation that the twist points
reflect some---as yet imperfectly understood---global aspect of the
QCD$_2$ string.

For the geometric interpretation of Yang-Mills theory to be viable,
each of the terms in the partition function expansion
must correspond to an actual covering of some sort.
For example, Kneser's theorem\cite{kneser}  states that
nonconstant continuous maps from a Riemann surface of genus $g$
(with $\Eu_g = 2-2g$) to a Riemann surface of genus $G$
(with $\Eu_G=2-2G$) do not exist if
\eq
               \Eu_g > r \Eu_G
\label{kneser}
\en
where $r$ is the degree of the map.
The same nonexistence constraint holds for
covering maps between nonorientable surfaces, with the
relation between Euler characteristic and genus now
given by $\Eu_q = 2-q$.
(This result follows from Kneser's theorem for the double covers of
the nonorientable surfaces.)
Remarkably, no terms arise in the Yang-Mills partition function
that violate these constraints.\cite{gross,nrs}

In this paper we will consider another restriction
on the terms of the Yang-Mills partition function.
It is a topological (rather than complex-analytic)
fact that the total number of branch points (counted with multiplicity)
must be even for any branched covering of an orientable or nonorientable
surface without boundary. To see this, consider,
for an $r$-fold branched covering with $n$
branch points $\{P_i\}$, the group homomorphism from
$\pi_1(\cM-\{P_i\})$ to $S_r$, the symmetric group on $r$ elements,
induced by the covering.
Each branch point is homotopically equivalent to a certain number,
$m_i$, of simple branch points (called the multiplicity of the branch point).
Therefore, we may consider the associated covering with
$\sum_i m_i$ simple branch points.
Let $\gamma$ be the element of $\pi_1(\cM-\{P_i\})$
that encloses all the branch points and separates them
from all handles and cross-cap locations (\ie, $\gamma$, considered
as an element of $\pi_1(\cM)$, is a
closed curve homotopic to the identity in $\pi_1(\cM)$).
With the surface constructed in the standard way as a
polygon with appropriate edge identifications,
it is clear that $\gamma$ is
\setcounter{footnote}{1}
homologous\footnote{We will be able to ignore the difference between
homology and homotopy here, because they differ by commutators;
the significance of this will be apparent in a moment.}
to the product of
generators (along the edge of the polygon) that appear
in the defining relation of $\pi_1(\cM)$.
If $\cM$ is orientable this defining relation is a product of
commutators. Since the group homomorphism
$\pi_1(\cM-\{P_i\}) \rightarrow S_r$ maps
commutator subgroup to commutator subgroup (as any group
homomorphism must), $\gamma$ is mapped to a commutator of permutations,
and any such permutation must have even
parity.
If $\cM$ is nonorientable the product of
generators to which $\gamma$ is homologous
is a product of squares of
elements, so that $\gamma$ is mapped to a product of squares
of permutations, and any such product must itself be an even permutation.
(Since $\gamma$ is an element of $\pi_1(\cM-\{P_i\})$ we have
not shown that $\gamma$ is mapped to the identity.)
Now deform $\gamma$ into a series of closed loops around each branch point.
As elements of $\pi_1(\cM-\{P_i\})$) each of the $\sum_i m_i$ loops
around a simple branch point is mapped to a transposition in $S_r$.
Since the product of these transpositions must equal
the {\em even} permutation that $\gamma$ is mapped to,
there must be an even number of transpositions, so that
$\sum_i m_i$, the total branch point multiplicity, must also be even.

It was pointed out in ref.~\dcite{nrs} that
there is a ${\ZZ}_2$-symmetry of the large-$N$ expansion associated
with Young tableau transposition that gives exactly
the cancellation of twist-point-independent terms
in the Yang-Mills partition function
required to ensure that we only need coverings with
an even number of simple branch points.
For $\cM$ a torus or Klein bottle (\ie, for $\Eu=0$),
twist points do not contribute to the partition function
and no more need be said.
When $\Eu \neq 0$, however,
there are cross-terms between simple branch points and
twist points, and it is not manifest {\it a priori}
that the topological constraint which allows only covers with an even number
of simple branch points is satisfied.
The failure of the simple branch point terms
to satisfy this constraint would put
the string interpretation of QCD$_2$ in jeopardy.

It is the purpose of this letter to lay that possibility to rest.
We show that the tableau transposition symmetry of the Yang-Mills partition
function on an arbitary orientable or nonorientable surface extends to
all (non-exponentially-suppressed) terms in the $1/N$ expansion.
This result guarantees that, if the twist points are
interpreted as multiple branch points, the total branch point multiplicity
is always even, to all orders in $1/N$.

The Yang-Mills partition function with arbitrary gauge group $G$
on a compact surface $\cM$ (with no boundary,
Euler characteristic $\Eu$, and area $A$) is given by
\eq
Z = \sum_{R\in \cO} (\sign_{\sR} \dim R)^{\Eu} \;
     \ex^{-\lambda A C_2(R)/2N} \; .
\label{partfn}
\en
If $\cM$ is orientable, then $\cO$ contains all irreducible
representations $R$ of $G$, but if $\cM$ is nonorientable,
then $\cO$ is restricted to self-conjugate representations.
For $\cM$ orientable, set $\sign_{\sR}=1$ for all $R$,
but for $\cM$ nonorientable,
the sign $\sign_{\sR}$ equals $+1 (-1)$ if there is a symmetric
(anti-symmetric) invariant in $ R \otimes R \to \CC$.
The dimension and quadratic Casimir of $R$ are denoted by
$\dim R$ and $C_2(R)$ respectively,
and  $\lambda = \gauge^2 N$ (where $\gauge$ is the gauge coupling constant).

Only representations with Casimirs much less than $N$ contribute
perturbatively to the $1/N$ expansion of the partition function.
For $\son$ and $\spn$, such representations
are in one-to-one correspondence with the Young tableaux composed
of $r$ cells with $ r \ll N $.
In particular, only tensor representations of $\son$ contribute
perturbatively.
Neglecting  nonperturbative effects in $1/N$,
the $\son$ and $\spn$ partition function becomes
\eq
Z  = \sum_{r}  \sum_{\RinY} (\sign^r \dim R)^{\Eu} \;
\ex^{-\lambda A C_2(R)/2N} \; ,
\label{partfcn}
\en
where $\cY_r$ denotes the set of Young tableaux with $r$ cells, and
$ \sign= 1$ for $\son$ and $-1$ for $\spn$.
Expression (\ref{partfcn}) is valid for
nonorientable as well as orientable surfaces,
since all the representations of $\son$ and $\spn$
in $\cY_r$ for $r \ll N$ are self-conjugate.
For $\sun$,  expression (\ref{partfcn}), with $\sign = 1$,
corresponds to either of two chiral sectors. (On a
nonorientable surface, the only $\sun$ representations that
contribute perturbatively are composite representations\cite{grotay} of
the form $R= S\bar{S}$, which all have $\sign_{\sR} =1$.)
The complete partition function is obtained by coupling the two
chiral sectors together.\cite{grotay}
The two sectors correspond to separate coverings,
connected by infinitesimal tubes in certain cases, and it
is conceivable that there could be an odd number
of branch points in each sector of such a coupled term.  In fact,
this does not occur and the number of branch points is even
in each sector.

We now cast (\ref{partfcn}) into a form
suitable for a $1/N$ expansion.
One may write\cite{twist} the character $\hchi_{\sR}^\G (U) $
of an arbitrary group element $U$
in the representation specified by the tableau $R$ of $G$
in terms of the new basis functions
\eq
\W_\sig^\G (U) =
\sum_{\RinY}  \hchi_{\sR} (\sigma) \hchi_{\sR}^\G (U),\qquad \sinS \;
\en
where $\hchi_{\sR} (\sigma) $ is the character of $\sig$
in the representation of the symmetric group $S_r$ that is also
denoted by the Young tableau $R$.
Orthonormality of the symmetric group characters gives
the inverse relation
\eq
\hchi_{\sR}^\G (U) = {1 \over r!}
\sum_{\sinS}  \hchi_{\sR} (\sig) \W_\sig^\G (U)\; , \qquad \RinY \; .
\en
Setting $U=1$, we obtain, for the dimension of $R$,
\eq
\dim R=   {1 \over r!}
\sum_{\sinS}  \hchi_{\sR} (\sig) \W_\sig^\G \; , \qquad
\W_\sig^\G \equiv \W_\sig^\G (1) \; .
\label{dimform}
\en
Explicit expressions for $\W_\sig^G$,
which will not be needed here,
are given in ref.~\dcite{twist} for $\sun$
and in ref.~\dcite{ram} for $\son$ and $\spn$.
In terms of the $S_r$ Frobenius algebra element
\eq
\Omega_{\G,r} = {1\over N^r} \sum_{\sinS} \W_\sig^\G \; \sig
\en
expression (\ref{dimform}) becomes
\eq
\dim R= {N^r \over r!} \hchi_{\sR} ( \Omega_{\G,r} ) \; .
\label{dimR}
\en
Since $\W_\sig^\G $ is a class function of $S_r$,
$\Omega_{\G,r}$ commutes with all the elements of $S_r$.
It is therefore a (non-zero) multiple of the identity matrix
in any representation $R$, by Schur's lemma, so that raising
it to any positive or negative power makes sense.
As a result,
\eq
\hchi_{\sR} (\Omega_{\G,r}^\Eu) =
   (d_{\sR})^{1-\Eu} \left[ \hchi_{\sR} (\Omega_{\G,r}) \right]^\Eu \; ,
\label{schur}
\en
where $d_{\sR}$ is the dimension of the symmetric group representation $R$.
With the coefficients $\om$ defined by
\eq
(\Omega_{\G,r})^\Eu = \sum_{\sinS}  \om \; \sig \; ,
\label{omdef}
\en
we arrive at
\eq
(\dim R)^\Eu = \left(N^r \over r! \right)^\Eu  (d_{\sR})^{\Eu-1}
\sum_{\sinS} \om \hchi_{\sR} (\sig) \; ,
\label{dimREu}
\en
by combining (\ref{dimR})--(\ref{omdef}). Note that
$\om$ is a power series in $1/N$.

The quadratic Casimir $C_2(R)$ is given by
\eq
C_2 (R) = fN \left[ r - U(r) + {T(R) \over N} \right]
\label{cas}
\en
where
\eq
\begin{array}{rcl}
f &=&\cases{ 1 & for $\sun$ and
               $\son$, \cr {1 \over 2} & for $\spn$, \cr}\\[0.3cm]
U(r) &=& \cases{  r^2/N^2 &  for $\sun$, \cr
		\sign r/N &  for $\son$ and  $\spn$, and \cr} \\[0.3cm]
T(R) &=& \ds \sum^n_{i=1} \, \ell_i (\ell_i + 1 - 2i)
     = \sum^{\col_1}_{i=1} \, \ell^2_i - \sum_{j=1}^{\ell_1} \col^2_j \; ,
\end{array}
\en
with $\ell_i$ ($\col_i$) denoting the row (column) lengths of the tableau $R$.
By using (\ref{dimREu}) and (\ref{cas}), we can write
the perturbative partition function (\ref{partfcn}) as
\eq
Z = \sum_{r} \left(\sign^r N^r \over r! \right)^{\Eu}  \;
\; \ex^{-\cA r/2} \; \ex^{\cA U(r)/2}  \;
\sum_{\sinS} \om \;  Z_{r,\sig}
\label{finpart}
\en
with
\eq
Z_{r,\sig} =
   \sum_{\RinY} \; (d_{\sR})^{\Eu-1} \hchi_{\sR} (\sig) \ex^{-\cA T(R)/2N}  \;
{}.
\en
Here $\cA = f \lambda A$ denotes the dimensionless area of the surface $\cM$.

In the geometric interpretation of (\ref{finpart}), the
sum over $r$ is interpreted as a sum over $r$-fold coverings
of $\cM$ by the string worldsheet. Each term is composed of
factors associated with particular features of the covering.
Each factor of $\cA T(R) /2N$ in a term indicates the presence of
a simple branch point, and each factor of $\cA U(r)/2$ corresponds to
the insertion of an infinitesimal handle for $\sun$,
or an infinitesimal cross-cap for $\son$ and $\spn$.
A major result of the work in refs.~\dcite{gross,minahan,grotay,twist,nrs}
is the demonstration that the leading term of
\eq
    \sum_{\sinS} \om \; Z_{r,\sig} = (d_{\sR})^{\Eu} \ex^{-\cA T(R)/2N} +
\ldots
\label{expand}
\en
has exactly the behavior expected
for a simple branch point; namely, that it behaves like
a transposition of two covering sheets.
The subleading terms indicated by the ellipsis are due to the presence
of twist points.
These terms correspond to a linear combination
(with coefficients $\om$) of coverings,
on each of which the sheets are permuted
according to the permutation $\sig$.
Each non-trivial permutation is equivalent to a product of
transpositions, which suggests the interpretation of a twist point
of given permutation type as a multiple branch point
homotopic to a number of simple branch points,
whose number is even (odd) if the permutation $\sig$ is even (odd).

We now show that the tableau transposition symmetry of the large-$N$ expansion
guarantees that the total branch point multiplicity is even, if
each $Z_{r,\sig}$ is taken to correspond to a covering with
multiple branch points. For each $\RinY$ with $r \ll N$,
there corresponds a transposed tableau $\tR$, also in $\cY_r$,
obtained by interchanging rows and columns ($\row_i(\tR) = \col_i(R)$).
Under this operation, the dimension and character of
the symmetric group representation $R$ obey\cite{littlewood}
\begin{eqnarray}
d_{{\sst \tR}} &=& d_{\sR}  \\
\hchi_{{\sst \tR}} (\sig) &=&
 \cases{ +\hchi_{\sR} (\sig)    & for $\sigma$ even\cr
         -\hchi_{\sR} (\sig)    & for $\sigma$ odd.}
\end{eqnarray}
The quantity $T(R)$ in (\ref{cas}) obeys\cite{dual,gross}
\eq
T(\tR) = - T(R) \; .
\en
Thus
\begin{eqnarray}
Z_{r,\sig}
& = & {\ts {1\over 2}}
\sum_{\RinY} \; \left[ (d_{\sR})^{\Eu-1} \hchi_{\sR} (\sig) \ex^{-\cA T(R)/2N}
+ (d_{{\sst \tR}})^{\Eu-1} \hchi_{{\sst \tR}} (\sig)
   \ex^{-\cA T(\tR)/2N} \right] \\
& = &
\sum_{\RinY} \; (d_{\sR})^{\Eu-1} \hchi_{\sR} (\sig)
\cases{  \cosh(-\cA T(R)/2N) & for $\sig$ even \cr
         \sinh(-\cA T(R)/2N) & for $\sig$ odd.}
\label{finalresult}
\end{eqnarray}
It is important to note that
the simple branch point terms $T(R)/2N$ do {\em not} all enter with
even powers. For gauge groups $\son$ and $\spn$ this is a
property of the complete partition function, but for $\sun$ it
is true of each chiral sector separately. Since the presence or absence
of infinitesimal tubes connecting the chiral sectors of the
$\sun$ Yang-Mills partition function is arbitrary,
coverings will appear with the chiral sectors disconnected
and with each
chiral sector containing an odd or even number of simple branch
points in an uncorrelated fashion.

Therefore, it is crucial for the geometric interpretation
that whenever the number of simple branch points
is odd (as in the expansion of $\sinh(-\cA T(R)/2N))$,
the twist point permutation $\sig$ is also odd.
Equation (\ref{finalresult}) shows that this is indeed the case.
(Note that in this case there must be at least one
cut connecting a simple branch point with a twist point.)
Similarly, whenever the number of simple branch points is even
(as in the expansion of $\cosh(-\cA T(R)/2N)),$
the twist point permutation $\sig$ is also even.

Thus, the total branch point multiplicity can only be even
if the twist points are interpreted as multiple branch points.
This interpretation therefore seems to be
essential for the consistency of the string interpretation.
{}From the vantage point of Yang-Mills theory
on both orientable and nonorientable surfaces,
the evenness of multiplicity constraint
is enforced by the transposition symmetry of the theory.

This conclusion makes the problem
of the lack of area dependence of the twist points more acute.
The factor of $\cA$
associated with each simple branch point term is interpreted
as the integration of the location of the ramification points
(those covering the branch points on $\cM$) over the worldsheet.
With the twist points taken to correspond to
multiple branch points, the lack of factors of $\cA$ in the twist point
terms is surprising,
especially in view of the necessary presence on some coverings
of cuts connecting mobile simple branch points with (apparently)
immobile multiple branch points.

\vspace{0.8cm}
\noindent {\bf Acknowledgement} \hspace{0.3cm}
We are grateful to D. Ruberman for the
branch-point-multiplicity evenness argument, a pedestrian
version of which appears in paragraph three.


\begin{thebibliography}{99.}
\bibitem{gross}  D. Gross,
                 \sl Nucl. Phys. \bf B400 \rm (1993) 161
\bibitem{minahan}J. Minahan,
                 \sl Phys. Rev. \bf D47 \rm (1993) 3430
\bibitem{grotay} D. Gross and W. Taylor,
                 \sl Nucl. Phys. \bf B400 \rm (1993) 181
\bibitem{twist}  D. Gross and W. Taylor,
                 \sl Nucl. Phys.  \bf B403 \rm (1993) 395
\bibitem{dk}  M. Douglas and V. Kazakov,
                 `Large $N$ Phase Transition
                  in Continuum ${\rm QCD}_2$,' \rm RU-93-17, LPTENS-93/20
                  (hepth{\smb}9305047)
\bibitem{nrs}  S. Naculich, H. Riggs, and H. Schnitzer,
		\sl Mod. Phys. Lett. \bf A8 \rm (1993) 2223
\bibitem{ram}  S. Ramgoolam,
		`Comment on Two-Dimensional O($N$) and Sp($N$) Yang-Mills
		Theories as String Theories,' \rm
		YCTP-P16-93 (hepth{\smb}9307085)
\bibitem{kos} I. Kostov, `Continuum QCD$_2$ in Terms of Discretized
                  Random Surfaces with Local Weights,' SPhT/93-050
                   (hepth{\smb}9306110)
\bibitem{recent} J. Minahan and A. Polychronakos, `Classical Solutions
                 for Two Dimensional QCD on the Sphere,' UVA-HET-93-08
                 (hepth{\smb}9309119); M. Caselle, A. D'Adda, L. Magnea,
                 and S. Panzeri, `Two-Dimensional QCD on the Sphere and on
                 the Cylinder,' DFTT 50/93 (hepth{\smb}9309107)
\bibitem{kneser}  H. Zieschang, E. Vogt, and H. Coldewey,
                 \it Surfaces and Planar Discontinuous Groups,
                 \rm Lecture Notes in Mathematics \bf 835 \rm
                 (Springer, New York, 1980)
\bibitem{littlewood} D. Littlewood,
		     \it The Theory of Group Characters
		     \rm (Oxford University Press, Oxford, 1950)
\bibitem{dual}  S. Naculich, H. Riggs, and H. Schnitzer,
		\sl Phys. Lett. \bf B246 \rm (1990) 417;
                S. Naculich and H. Schnitzer,
		\sl Nucl. Phys. \bf B347 \rm (1990) 687
\end{thebibliography}
\end{document}